\documentclass[twocolumn]{revtex4}
\usepackage[dvips]{graphicx, color}
\usepackage{amsmath,amssymb}
\usepackage{pspicture, graphpap}

\DeclareMathAlphabet\mathbb{U}{fplmbb}{m}{n}

\newcommand\ltap{\
  \raise.3ex\hbox{$<$\kern-.75em\lower1ex\hbox{$\sim$}}\ } 
\newcommand\gtap{\
  \raise.3ex\hbox{$>$\kern-.75em\lower1ex\hbox{$\sim$}}\ } 
\newcommand\simge{\mathrel{%
   \rlap{\raise 0.511ex \hbox{$>$}}{\lower 0.511ex \hbox{$\sim$}}}}
\newcommand\simle{\mathrel{
   \rlap{\raise 0.511ex \hbox{$<$}}{\lower 0.511ex \hbox{$\sim$}}}}

\newcommand{\slashchar}[1]%
        {\kern .25em\raise.18ex\hbox{$/$}\kern-.75em #1}
\def\lsim{\mathrel{\raise.3ex\hbox{$<$\kern-.75em\lower1ex\hbox{$\sim$}}}}
\def\gsim{\mathrel{\raise.3ex\hbox{$>$\kern-.75em\lower1ex\hbox{$\sim$}}}}
\newcommand\CL{{\cal L}}

\newcommand\CO{{\cal O}}

\newcommand\CZ{{\cal Z}}
\newcommand\be{\begin{equation}} 
\newcommand\ee{\end{equation}} 
\newcommand\bea{\begin{eqnarray}}
\newcommand\eea{\end{eqnarray}}
\newcommand\ba{\begin{array}}
\newcommand\ea{\end{array}}

\newcommand\kev{{\rm keV}}

\newcommand\gev{{\rm GeV}}
\newcommand\tev{{\rm TeV}}

\newcommand\jet{{\rm jet}}
\newcommand\jets{{\rm jets}}

\begin{document}

\title{HIGGS CASCADE DECAYS TO $\gamma\gamma + \jet\ \jet$ AT THE LHC} 
\author{Adam Martin}
\email{adam.martin@yale.edu}
 \affiliation{Department of Physics, Sloane
 Laboratory, Yale University, New Haven CT 06520}
 \date{\today}
\vspace{-1.0in}
\begin{abstract}
Extra light electroweak singlets can dramatically alter Higgs decays by introducing additional
decay modes, $h\rightarrow aa$. In scenarios where cascade decays
$h\rightarrow aa\rightarrow 4X, X\ne b,\bar b$ dominate, the Higgs
will escape conventional searches and may be as light as $82\ \gev$. In this paper we investigate the
discovery potential of the mode $h\rightarrow aa\rightarrow 2\gamma 2g$ through
direct ($pp\rightarrow h$) and associated ($pp \rightarrow W^{\pm}h$) Higgs production at the LHC. Our search covers all
kinematically allowed singlet masses for $\sim 80\ \gev \le m_h < 160\
\gev$ and assumes an integrated luminosity $\CL = 300\ {\rm
  fb}^{-1}$.  We find associated production, despite a smaller production
cross section, to be the better mode. A branching ratio
$BR(h\rightarrow 2\gamma 2g) \cong
0.04$ is sufficient for discovery in the bulk of our search
window. Given the same luminosity and branching ratio $0.04$, direct
detection fails to discover a Higgs anywhere in our search window.
Discovery in the limited region  $m_h > 120\ \gev, m_a \sim 25\ \gev$ is
possible with direct production when
the branching ratio is $\simge 0.06$.

%  Given the same luminosity and
% branching ratio $0.04$, we find discovery through direct production only when
% $m_h \simge 150\ \gev$ and $m_a \sim 20\ \gev$. Substantially
% expanding the direct detection
% discover region requires branching ratios $> 0.1$.

% Branching ratios $\sim 0.2$ are sufficient for Higgs discovery for only a
% small subset of $m_h, m_a$ values, and branching ratios of order $\sim
% 0.1$ are necessary to cover 

% Despite a smaller production
% cross section, we find associated production to be the better
% detection mode. Given an integrated luminosity of $300\ fb^{-1}$, a
% branching ratio $BR(h\rightarrow 2\gamma 2h) \cong
% 0.02$ is sufficient for discovery in the bulk of our search
% window. 

\end{abstract}
  \maketitle

\section{Introduction and Motivation:}

% Higgs decay to a pair of pseudoscalars can easily dominate over SM
% higgs decays, especially when $m_h \le 160\ \gev$. If the subsequent pseudoscalars do not decay to $b$ quarks, then these
% higgs would escape the conventional detection techniques. Also, the higgs in
% this scenario may be much lighter than the usual $114.4\ \gev$ LEP bound. The
% only relevant 

% Here and throughout, fine tuning is measured by the
% specificity of the NMSSM parameters needed to achieve the spectrum.

Electroweak singlet fields are common in extensions of the Standard
Model (SM). Some models require singlets for theoretical reasons, while in
other models they can be simply tacked onto the existing particle
content. Regardless of whether extra singlets are required, Higgs decay
in their presence can be dramatically different
than in the SM~\cite{O'Connell:2006wi, Bahat-Treidel:2006kx,
  Barger:2006rd, Barger:2006dh,Barger:2006sk,Arhrib:2006cv}. First, new decay modes $h\rightarrow aa$, where $a$ is the
electroweak singlet, open up, suppressing SM branching fractions. Second, assuming
that $a$ decays to SM fields, the Higgs signal becomes a cascade decay $h \rightarrow
aa \rightarrow X$, where $X$ contains four or more SM fields. The
resulting final state looks nothing like the final state of a SM Higgs. Depending on the mass of the Higgs and on the number and type
of SM fields the singlets decay into, these cascade decays may avoid
conventional detection techniques and even allow for lower mass
Higgses. Given the ease which singlets can accompany beyond-the-SM
physics and the dramatic changes they can lead to in the the Higgs signal, it is important to
explore the discovery potential of the LHC in such scenarios.

In this paper we examine Higgses which cascade decays into
four visible, light SM particles.  We focus on this possibility because, while the LEP II bounds on exotic Higgs decay such as
$h\rightarrow bb\bar b \bar b$ and $h\rightarrow {\rm invisible}$ are
similar to the LEP II standard model Higgs $114.4\ \gev$
bound~\cite{LEPEWWG, Barate:2003sz, Schael:2006cr}, a Higgs which
decays primarily $h \rightarrow aa
\rightarrow 4X, X\ne b, \bar b$ may be much
lighter. The only pertinent bound comes from the OPAL decay-independent
 study~\cite{Abbiendi:2002qp} which requires $m_h \ge 82\
 \gev$.  Models which allow a light Higgs mass are particularly
 interesting given that indirect constraints from
 precision electroweak experiments also favor a light Higgs $m_h =
 84^{+32}_{-25}\ \gev$~\cite{Erler:2007sc}. Explicit models with this Higgs decay
 structure were recently studied in Ref.~\cite{Chang:2005ht} within
 extensions of the NMSSM. Of the remaining possibilities for $X$, we
will not discuss $a \rightarrow \tau^+ \tau^-$ because it requires
$\sim$ percent level fine tuning (to get $2 m_{\tau} < m_a < 2 m_b$)
within the class of models in~\cite{Chang:2005ht}. Instead, we focus on the decays $a \rightarrow \gamma \gamma,
gg$.

In order for the $gg, \gamma\gamma$ decay modes to dominate without
making $a$ very light, $a\rightarrow {\rm fermion}$ decays must be suppressed. A simple and natural way this
can happen is if $a$ is odd under some $\CZ_2$ symmetry, while all SM
fermions are even. If the $\CZ_2$ is not broken, then the $a$ will be
stable and this mode is subject to the LEP $h\rightarrow {\rm
  invisible}$ constraint $m_h > 114.0\
\gev$~\cite{unknown:2001xz}. However, if $\CZ_2$ is broken by only the
coupling of $a$ to new, heavy
vector-like fermions $Q_i$, then $a$ decays to SM fermions are
forbidden and
$a$ decays solely to photons or gluons through $Q_i$ loops. We will take the
$\CZ_2$ symmetry to be CP, and thus the electroweak singlet $a$ is a
CP-odd pseudoscalar.

The dominant Higgs decay mode in this scenario is $h \rightarrow aa
\rightarrow 4 g$. It would be swamped by QCD background at the LHC. The cleanest decay mode, $h
\rightarrow 2 a \rightarrow 4 \gamma$, was investigated in
Ref.~\cite{Chang:2006bw}. It suffers from a very small branching
ratio and therefore requires a lot of luminosity. Even with $300\ {\rm
  fb}^{-1}$ of luminosity, branching ratios typical of
Ref.~\cite{Chang:2005ht} $(\sim 10^{-5})$ are
too small to be discovered in the majority of $m_h, m_a$ parameter space.
 Our
goal is to explore the discovery potential of the LHC in
the mixed decay mode $h \rightarrow 2\gamma\ 2g$. This mode has the
best of both worlds - a higher branching ratio than $h \rightarrow 4
\gamma$ and less background than $h \rightarrow 4g$. Previously
investigations of this decay mode were very preliminary and focused
on light $a$ and $m_h \ge 100\
\gev$~\cite{Dobrescu:2000jt,Landsberg:2000ht}. We study both
direct $pp\rightarrow h$ and associated $pp\rightarrow W^{\pm}h$
production.

We find direct production suffers from large irreducible backgrounds,
primarily $\gamma\gamma + \jets$. To reduce the background, 
we impose additional cuts on the angular separation of the
pseudoscalar decay products. These extra cuts force us into a smaller subset of $m_h, m_a$ space. The particular
cuts we choose restrict us to scenarios $15\ \gev < m_a < 35\
\gev$. Within that $m_a$ band and assuming $\CL = 300\ {\rm fb}^{-1}$, we
find branching ratios $BR(h\rightarrow 2\gamma 2g) \sim
0.06$ are sufficient only for heavy
Higgs ($> 120\ \gev$) detection. Higgses lighter than $90\ \gev$ remain undetectable
unless the branching ratio is $> 0.2$.

%  It may be possible to
% enhance the significance in this channel through a two-dimensional
% Higgs-pseudoscalar search.

%  but not enough for discovery in the majority of $m_h, m_a$ space. However, once the mass of
% the pseudoscalar is known, most of the background can be discarded and
% the significance greatly improves. We incorporate the effect of the
% pseudoscalar discovery through a
% simultaneous pseudoscalar - Higgs search.

Associated production is a productive alternative. The
production cross section is smaller, thus high luminosity is still
necessary. As in direct production, angular cuts increase the significance of the light-$m_a$
region of parameter space. Given $300\ {\rm fb}^{-1}$, we find the Higgs
can be discovered with branching ratios $BR(h\rightarrow 2\gamma 2g) = 0.04
$ in the majority of $m_h, m_a$ space. For $m_a \sim  30\ \gev$
 and $m_h$ between $100\ \gev$ and $120\ \gev$, the detection
 prospects are even better.
 Additionally, the regions not sufficiently probed
by associated production are exactly the regions with the highest
significance in $h\rightarrow aa\rightarrow
4\gamma$~\cite{Chang:2006bw}.

The setup of this paper is as follows: In section II we review the
operators and parameters necessary for $h\rightarrow 2\gamma 2g$
decays. In section III we describe the signal and background simulation
procedure for direct production. Associated production is described in section IV. Results are presented at the end of each
production mode section. Conclusions are given in section V.

%  In particular, we want to study
% \begin{itemize}
% \item What regions of $m_a - m_h$ space are probed at a given luminosity?
  
% \item Can we say anything about
%   $m_h$ or will we just see an excess in some channel?
  
% \item How does this mode complement the $h \rightarrow 4 \gamma$ mode
%   of~{\cite{Chang:2006bw}}?
% \end{itemize}

\section{Interactions and Branching Ratios}

In this section we introduce the interactions and parameters necessary
for cascade Higgs decays $h\rightarrow aa \rightarrow 2\gamma
2g$. Although presumably this scenario is embedded into some larger
 new physics model, the only fields we will need are the $SU(2)_w$ doublet SM Higgs $H$,
the new electroweak singlet $a$, and one or more vector-like fermions $Q_i$.

Because of CP and $SU(2)_w$
invariance, trilinear $H - a$ operators are forbidden so the lowest dimension
operator we can write down which includes both $H$ and $a$ is
\be
\label{eq:haa}
 \frac{\kappa}{2\sqrt
  2} (H^{\dag} H)(a^2).
\ee
After EWSB, this
operator contains the interaction $\frac{\kappa~v}{\sqrt 2}(h a^2) $, where $h$ is the
physical Higgs boson and $v = 246\ \gev$. This allows $h\rightarrow
2a$. The strength of the $h \rightarrow aa$ decay mode depends on
$\kappa$ and the mass of the Higgs. The lower limit of our Higgs search is the
OPAL bound, $\sim 82\ \gev$, and we set an upper limit of $160\
\gev$. Above $m_h \sim\ 160\ \gev$, the $h\rightarrow W^+W^-$ mode
opens and we expect it to dominate. Within this range
of Higgs masses, small $\kappa \ll 1$ are sufficient for
$h\rightarrow aa$ to be the dominant mode. For example, $\frac{\kappa v}{\sqrt 2} \gtrapprox 5\ \gev $
for a $100\ \gev$ Higgs is sufficient. 

The pseudoscalar $a$ decays because of its coupling to heavy
vector-like fermions,
\be
\lambda~a~(\bar Q_i \gamma_5 Q_i).
\ee
Upon integrating out the heavy fermions $Q_i$, this interaction generates
effective $a\gamma\gamma$ and $a gg$ operators~\cite{Dobrescu:2000jt,Chang:2005ht}
\be 
\frac{\lambda}{8\sqrt 2\pi M_Q}a\Big(
(b_3~\alpha_3)G^a_{\mu\nu}\tilde G^{a\mu\nu} +
(b_{em}~\alpha) F_{\mu\nu}\tilde F^{\mu\nu} \Big),
\ee
where $\lambda$ is the coupling of the pseudoscalar
to the vector-like fermion of mass $M_Q$. The interactions are
proportional to the $SU(3), U(1)_{em}$ gauge couplings $\alpha_i$ and to the contribution of the
vector-like fermions to the corresponding beta function $b_{3}, b_{em}$.
From these interactions, we can derive the decay rates $a\rightarrow \gamma\gamma, gg$: 
\be
{\Gamma}_i = {\frac{9{\lambda}^2 b^2_i {\alpha}^2_i}{1024 {\pi}^3
    M^2_Q}}m^3_a N_V,
\ee
where $N_V$ counts the number of gauge bosons (1 photon, 8 gluons).

 For a given Higgs mass, we consider all kinematically allowed pseudoscalar
masses.  This doesn't conflict with our constraint on $\kappa$, since
nothing forbids additional pseudoscalar mass terms $\mu^2 (a a)$.

 % Although the operator~(\ref{eq:haa}) contributes
% $\CO(\kappa v^2)$ to the mass of $a$, the term $m^2_a (a a)$ is
% also allowed and hence the mass of $a$ is not constrained by $\kappa$.
%  We are free to consider all kinematically allowed psuedoscalars.

The relevant parameter for this study is the branching ratio
$\text{\textrm{BR}} ( h \rightarrow 2 \gamma 2 g )$. Assuming the Higgs
always decays into a pair of psuedoscalars,
\be 
BR(h\rightarrow 4\gamma) \cong (0.5\times BR(h\rightarrow
2\gamma 2g))^2.
\label{eq:compare}
\ee
we can make a rough comparison between our results and the $h\rightarrow 4\gamma$ results of
Ref.~\cite{Chang:2006bw}. This branching ratio is
determined solely by the quantity and quantum numbers of the $Q_i$.
\be
\label{eq:BRhaa}
{\frac{{\Gamma}_{{\gamma} {\gamma}}}{{\Gamma}_{gg}}} = {\frac{b^2_1
    {\alpha}^2}{8~b^2_3 {\alpha}_3^2}},
\ee
To get an idea of a typical branching ratio, consider the case where $Q$ is a single $5$ of $SU ( 5 )$
($d^c$ plus $L$). Then $b_3 = \frac{2}{3}, b_1 = \frac{2}{3} ( 1 - \frac{1}{3} )
= \frac{4}{9}$, and the ratio is $3.8 \times 10^{- 3}$,  which makes $BR ( h
\rightarrow 2 g 2 \gamma ) = 7.6 \times 10^{- 3}$. We will use this value as
a benchmark point. The branching ratio to photons can be enhanced if we couple $a$
to additional color singlet matter (e.g. higgsinos). For an
approximate upper bound on the cross section, we follow
Ref.~\cite{Chang:2006bw} and assume that LEP
can place an effective limit of $BR(h\rightarrow 4\gamma) \simle
\CO(10^{-3})$ for $m_h < 120\ \gev$. In our calculations we will not
consider branching ratios higher than $0.2$ at any $m_h$.

Although the branching ratio
(\ref{eq:BRhaa}) is independent of both the heavy fermion mass $M_Q$
and its coupling $\lambda$ to $a$, the ratio of these parameters $\frac{\lambda}{M_Q}$ does
set the scale for the total width of the
pseudoscalar. For $\lambda \sim 1$ and $M_Q \sim 1\ \tev$, a typical
width is $\CO(\kev)$. While it is possible to choose parameters such that
$\frac{\lambda}{M_Q} < 1.5 \times 10^{-3}\ \tev^{-1}$ and the $a$ decay
vertices are displaced from the original vertex by an
experimentally detectable amount~\cite{Chang:2006bw}, we
will ignore that possibility here.

%  Though displaced
% vertices may help distinguish this Higgs signal from the background, 

%  A short pseudosclar lifetime $c\tau \ll $ is possible without extreme choices for either $M_Q$ or $\lambda$. In
% particular, 

% Although the branching ratio (\ref{eq:BRhaa}) does not depend explicitly on the mass of
% the heavy fermions $M_Q$ or its coupling $\lambda$ to the
% pseudoscalar, . 

%  $M_Q$ can be constrained if we require the $a$ to decay within
% the detector. If $a$ does not decay within the detector we must comply with
% the stringent invisible Higgs decay bound $m_h \ge 114 \gev$. We
% assume that $a$ decays within the detector throughout this paper.
 
\section{Direct Production}

%\subsection{Direct Production:}

The dominant method of Higgs production at the LHC is gluon fusion.
Gluon fusion is a loop level process and is therefore sensitive to
new physics. While it is certainly possible that the Standard Model
extensions (singlets, vector-like colored fermions) which lead to $h
\rightarrow aa$, $a \rightarrow \gamma \gamma, gg$ will effect Higgs
production, the net effect is hard to estimate and depends on which
model (SUSY, little Higgs, etc.) we embed the $h\rightarrow aa$ scenario into. We therefore use the SM Higgs
production cross section throughout this work.

%  In addition to loops of
% additional weak-scale colored particles
% (squarks, additional vectorlike quarks, etc.)
% higher dimensional operators $\propto h GG$ may also be
% important~\cite{Manohar:2006gz}.

% To demonstrate this, assume we are working within the context of the NMSSM.
% Then $a$ is part of a new singlet superfield which also contains a real
% scalar $s$. This singlet mixes with the physical Higgs boson once EWSB
% occurs, forming mass eigenstates $\widetilde{s_{}}$ and $\tilde{h}$. The
% coupling of $s$ to the heavy vector-like quark is required by SUSY and leads
% to effective $sgg$ and $s\gamma \gamma$ vertices. This yields
% contributions to the gluon fusion amplitude proportional to
% \be
% {\frac{{\lambda}}{M_Q}}\sin{\theta}
% \ee
% where theta is the $s - h$ mixing
% angle.

%  However, we have to combine this effect with the contribution to $gg
% \rightarrow$ from squark loops. The squark contribution can have either sign
% and depends sensitively on the squark masses and the amount of
% mixing.

We generate the signal $h \rightarrow aa$, $a \rightarrow 2
g, 2 \gamma$ events using PYTHIA 4.0~\cite{Sjostrand:2003wg}. Within
PYTHIA, the matrix-element level events are parton showered and
hadronized. In the showering process PYTHIA adds initial/final state radiation and multiple
interactions. Signal events were generated using the CTEQ5L~\cite{Lai:1999wy} parton
distribution functions.

%  all objects with transverse energy $E_T >
% E_{T,thresh}$ for a given event. After the calorimeter clusters are
% formed, objects are reconstructed using
% a cone of size $R_{rec}$.

 Once hadronized, all events are run
through the detector simulator PGS 4.0~\cite{PGS} which incorporates detector efficiencies and smearing
effects. In PGS, the calorimeter is divided into segments in
$\eta -\phi$ space, where $\eta$ is the pseudorapidity and $\phi$ is the axial angle. Within
the $(\eta,\phi)$ grid, PGS forms calorimeter clusters around all
cells with transverse energy greater than $E_{T, thresh}$. The clusters are then reconstructed into physical particles using a cone algorithm. We use a segmentation $\Delta
\eta \times \Delta \phi = 0.1\ \times\ 0.1$, an $E_T$
threshold of $5.0\ \gev$, and a cone of size $R_{rec} = 0.4$. The other important
detector parameters for our analysis are the jet and photon energy resolutions, and
the photon reconstruction procedure. We use energy resolutions
\be
 \frac{\delta E_{jet}}{E_{jet}} = \frac{0.8}{\sqrt E_{jet}(\gev)}~~,~~
 \frac{\delta E_{\gamma}}{E_{\gamma}} = \frac{0.1}{\sqrt{E_{\gamma}(\gev)}}
 + 0.007 
\ee
throughout. To reconstruct photons, we use the default PGS 
procedure: In order for a final state object to be considered a
photon, it must have $p_T > 5.0 \ \gev$, no track, and a ratio of hadronic
calorimeter $E_T$ to electromagnetic calorimeter $E_T$ $<
0.125$. In addition photons are required to be isolated, meaning the total $E_T$ in the $3 \times 3$ ring of calorimeter segments around the photon must
be less than $10\%$ of the photon's $E_T$, and the total $p_T$ from
tracks within a cone $\Delta R = 0.4$ of the photon must be less that
$5.0\ \gev$.

We include a $K$ factor of $2.0$ to
account for higher order contributions to the signal. With this factor we recover the NNLO SM
production cross section of Ref.~\cite{Djouadi:2005gi}.
\\

Initial Cuts:
\begin{itemize}
\item 2 $\gamma$'s with $p_T > 20.0$ and $| \eta | < 2.5$
  
\item 2+ $\jets$ with $p_T > 20$ and $| \eta | < 2.5$
  
\item $\Delta R_{jj} \ge 0.4$ (between any jet pairs)
  
\item $\Delta R_{\gamma \gamma} \ge 0.4$
  
\item $\Delta R_{\jet - \gamma} \ge 0.4$
    
%  Pileup is included in ATLFAST as an additional $p_T$-dependent
% smearing parameter for all objects.

% \item $|m_{jj} - m_{\gamma\gamma}| < 50\
%   \gev$

\end{itemize}
% In the last cut, $m_{\gamma\gamma}$ is the invariant
% mass of the photon pair and $m_{jj}$ is the invariant mass of the leading two
% jets.

% (jet pair with the invariant mass closest to ... ).

The efficiency for the signal under the isolation cuts varies between $3\% - 15\%$. It 
increases as $m_h$ and $m_a$ increase, though the dependence on $m_a$ is
stronger. Scenarios with $m_a \lessapprox 10\ \gev$ are excluded by
these cuts. For
such light pseudoscalars, the subsequent decay products (gluons or
photons) are too collinear and cannot pass the isolation cut $\Delta R =
0.4$.

%  We see
% no significant changes in the signal efficiency if we
% switch to stronger $\Delta R_{\gamma\gamma}, \Delta R_{j-\gamma}
% > 0.7$ isolation cuts. However a stricter $\Delta R_{jj} > 0.7$ cut does
% cause the signal in the low$-m_a$ region of parameter space to
% deteriorate as the individual $\gamma$s or jets from pseudoscalar
% decay cannot be resolved. In fact, scenarios with $m_a
% \lessapprox 10\ \gev$ are already excluded by the $\Delta R > 0.4$
% isolation cut. 

% To show the dependence of our search on the isolation
% cuts, we show all results for both $\Delta R$ values.

\subsection{Background}

The backgrounds, in order of importance are:
\begin{itemize}
\item diphoton + jets: $\gamma \gamma + N\ \jets, N = 0, 1, ..2$
  
\item photon + jets: $\gamma + N\ \jets, N = 0, ..3$
  
\item QCD multijets: $N\ \jets, N = 2, 3, 4$
\end{itemize}
% \item Although the our final state only requires 4 partons we consider
%   backgrounds with $> 4$ in an effort to conservatively estimate the
%   background.

All backgrounds were initially simulated with ALPGEN
v2.11~\cite{Mangano:2002ea}, showered and hadronized in PYTHIA, then run
through PGS. The parton distribution set CTEQ5L was used for all backgrounds. ALPGEN
contains various options for the factorization scale. We used the
choice $Q_{fac}^2 = \sum_i p_{T, \gamma_i}^2 + p_{T, j_i}^2$, where $i$ runs over
all jets and photons in the matrix element. ALPGEN imposes
generator level cuts on all events, which we take to be softer ($p_T >
15\ \gev, |\eta| < 3.5$) than
the isolation cuts to avoid losing any background.

One immediate concern when patching matrix element generators (ALPGEN) with
showering Monte Carlos (PYTHIA) in multijet events is jet overcounting. The current version of
ALPGEN employs the MLM jet-parton matching scheme~{\cite{Catani:2001cc}} to ameliorate this
problem. For a given background $X + N\
\jets$, we generate events using the MLM
 procedure for the subleading jet multiplicity processes, $ + 0\ \jets$ up to
$+ (N-1) \jets$. In order to capture the inclusive $ + \jets$ background, events
for the highest jet multiplicity process are generated without the matching.

In the latter two backgrounds, at least one jet fakes a photon. The
QCD fake rate is $p_T$ dependent.  For $20\ \gev <  p_T < 40\ \gev$ the jet rate
decreases approximately linearly, while above $p_T = 40\ \gev$ the fake rate
is nearly constant. In practice we use the same function as Ref.~\cite{Chang:2006bw}
\be
P({\jet}
{\rightarrow} {\gamma}) = 1.0/\textrm{min}(3067, -1333+110
p_T/{\gev}),
\ee
which was obtained by fitting the fake-rate plot in the ATLAS
TDR~{\cite{:1999fq}}. With this rate we find the single fake
background constitutes roughly $20\%$ of the background, while the two-fake
background is small, $\simle 1\%$.

%  For simplicity we will use
% the flat rate $P({\jet}
% {\rightarrow} {\gamma}) = 1.0/3000$, which is consistent with the
% ALTAS TDR~{\cite{:1999fq} for $p_T > 40\ \gev$. This value may be too
%   optimistic for lower $p_T$ jets, however, the jet-fake
%   backgrounds constitue only a few percent of the total background, so
%   we do not expect a change in this rate to have a very significant effect.

%  It is approximately the same as the fake-rate
% used in

%  however it is consistent with the current fake rate after full
% reconstruction (cite.).

% Rather than use a flat $K$ factor, we incorporate higher
% order QCD effects for the background by explicitly
% including higher order processes, e.x. $pp \rightarrow \gamma\gamma
% + 3\ \rm{hard}\ \jets$.

To include NLO effects we use a $K$ factor of $1.5$ for $\gamma + \jets$~\cite{unknown:1999fr}. For $\gamma\gamma +
\jets$, we take $K = 1.5$ for $\gamma\gamma + 0\ \jets$ and $K = 1.2$
for $\gamma \gamma + 1^+\ \jets$. The diphoton $K$ factors were chosen to best match to the LO
and NLO $pp
\rightarrow,\ \gamma\gamma + X$
results in Ref.~\cite{unknown:1999fr, Baer:1990ra, Balazs:1998bm}. Since the
QCD-fake background is so small, we do include a $K$ factor for the $N_{jet}$ processes.

%  Separate $K$ factor for each additional jet process
% are applied according t(cite.).

% ii.) Using the same $K = 2.0 $ factor
% that we use for the signal. To be conservative we take the
% larger result.

%  in which a scale $Q^2_0$ in introduced.
% Parton splittings above $y_{ini} = \frac{Q^2_0}{Q^2}$, where $Q^2$ is
% the energy of the initial hard process, are treated with the full matrix element
% and dressed with Sudakov form factors on each external leg. Splittings below
% $y_{ini}$ are treated by the parton showering algorithm. The leading order dependence on
% $y_{ini}$ cancels once all contributions to a particular jet configuration
% are included.

% Imposing the minimal cuts, the background $M_{\gamma \gamma}$
% distribution is approximately linear up to $~ 40\
% \gev$. Past $40\ \gev$ the slope increases and the shape becomes more
% gaussian with a peak around $60\ \gev$. The background $M_{jj \gamma \gamma}$
% distribution has an approximately constant positive slope between $\sim 100
% - 140\ \gev$. Above $140\ \gev$ the distribution starts to turn over, while
% below $100\ \gev$ $M_{jj \gamma \gamma}$ slowly tails off down to $\sim 60\
% \gev$.

\subsection{Reconstruction}

The first step in reconstructing the Higgs is to determine the invariant mass of the two photons, $M_{\gamma
   \gamma}$, for all events which pass the cuts. These photons are
 combined with the two leading $p_T$ jets to form the total mass of the $2 j 2
 \gamma$ system, $M_{jj \gamma \gamma}$. Our region of interest is $M_{jj
   \gamma \gamma} < 160\ \gev$ (and therefore $M_{\gamma\gamma} \le 80\
 \gev$). 
 
%whose invariant mass is closest to $M_{\gamma \gamma}$  ( CLOSEST INVARIANT MASS OR LEADING JETS?)
% Additional cuts:
% \begin{itemize}
   
 Reconstructing $M_{jj\gamma\gamma}$, the signal has a large
 tail containing events where the wrong jet pair was chosen or
 initial/final state radiation has ruined the mass peak. This tail, along with the
 background, are effectively suppressed by imposing additional
 angular cuts. The most efficient variables to cut on are the azimuthal separation of
the pseudoscalar decay products $\Delta
 \phi_{jj}, \Delta \phi_{\gamma\gamma}$, and the difference between the invariant mass of the two
photons and the invariant mass of the two leading jets, $\Delta M \equiv |m_{jj} - m_{\gamma\gamma}|$.
These three variables are shown below in fig.~(\ref{fig:cuts}) for the
background and an example $h \rightarrow 2\gamma 2g$ signal.

 % The most efficient variables to cut on are $\Delta
%  \phi_{jj}, \Delta \phi_{\gamma\gamma},\ {\rm and}\ 
% Delta M \equiv |M_{jj} - M_{\gamma\gamma}|$. 

 \begin{figure}[!h]
   \includegraphics[width=2.5in,height=2.25in]{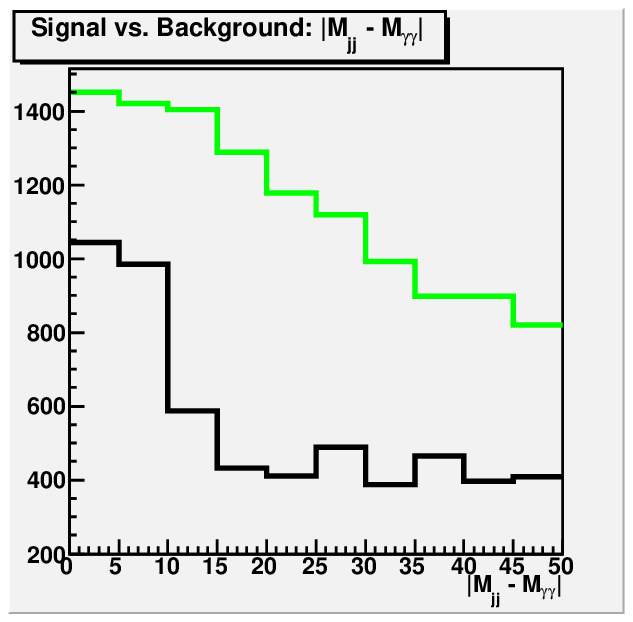}
   \includegraphics[width=2.5in,height=2.25in]{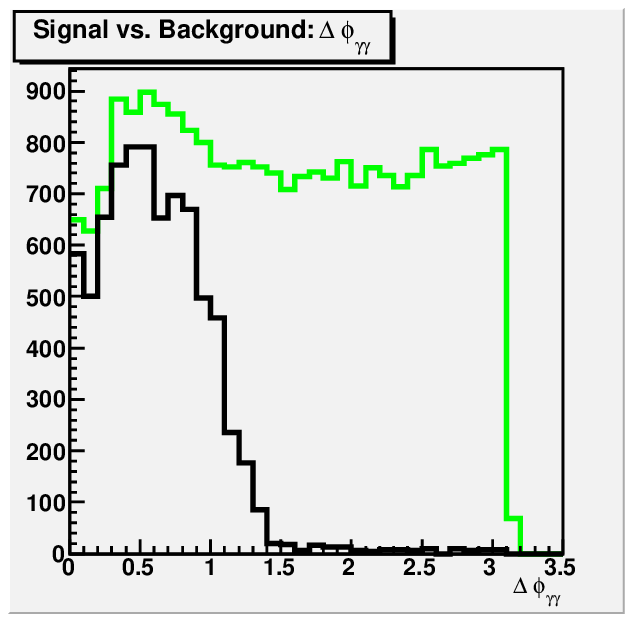}
   \includegraphics[width=2.5in,height=2.25in]{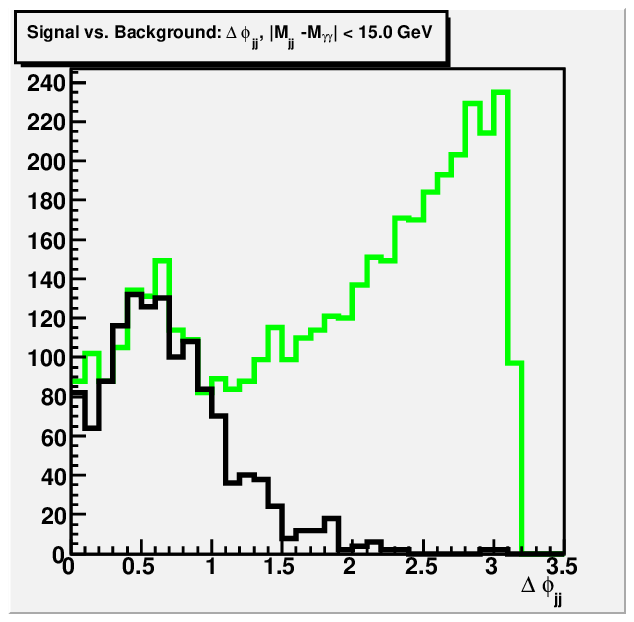}
   \caption{$\Delta M, \Delta \phi_{jj},\ {\rm and}\ \Delta
     \phi_{\gamma\gamma}$ for the signal (black), and background
     (gray, green online). The background distribution was constructed with a 50,000
   event sample. The signal assumes $m_h = 120\ \gev$, $m_a = 30\
   \gev$, and has been scaled up so the distribution shapes
   shapes can easily be compared. The $\Delta \phi_{jj}$ distribution
   was created after imposing a $\Delta M < 15.0\ \gev$ cut.}
   \label{fig:cuts}
 \end{figure}

 The signal $\Delta M$ distribution does not change much as $m_h$ and
 $m_a$ are varied. From fig.~(\ref{fig:cuts}) we see that a cut value
 of $\Delta M \simle 15\ \gev$ is most effective. The optimal cut
 values for $\Delta \phi_{jj}$ and $\Delta \phi_{\gamma\gamma}$ depend
 on the mass of the pseudoscalar. The lighter the pseudoscalar, the earlier the $\Delta \phi$ distribution
 peaks, and more background can be excluded. This can be understood
 from the decay kinematics in the following way.  When the Higgs decays into much lighter
 pseudoscalars, the $a$ will be produced with substantial kinetic
 energy. Their subsequent decay products (gluons or photons) are
 predominantly collinear. The higher the velocity, the smaller the
 average angular separation between the decay products. The most
 collinear events are removed by our isolation cut, leaving behind a
 peaked $\Delta \phi$ distribution. 
 In contrast, the background photons (or jets) are emitted primarily back-to-back, where $\Delta
 \phi_{\gamma \gamma}, \Delta \phi_{jj} \approx \pi$. Although a smaller
 $\Delta \phi$ implies a smaller $\Delta R$, we find the $\Delta
 \phi$ cuts to be more efficient. 

 % Cuts on
 % the maximal value for $\Delta R_{jj}$, $\Delta \phi_{\gamma \gamma}$, or
 % $\Delta \phi_{jj}$ have analogous results. The latter two may be more
 % practical because we have not already imposed a minimum cut.\\
 
 % This cut also reduces the tail (from combinatorics? ) of the signal
 % distribution which makes the peak more visible.\\
 
 % To further improve the significance, a cut $\text{\textrm{max}} \{
 % \Delta R_{\gamma \gamma} \} \lessapprox 1.0$ is effective.
 
 This angular cut is less effective for heavy $a$ which are produced with
 little kinetic energy. This is offset somewhat by the fact that heavier $m_a$ more efficiently pass
 the $p_T > 20.0$ cut. Since the kinematics of the
 signal is least like the background for light $m_a$, we will focus on
 cuts that enhance the significance of that region.

 Two variables which often show up in SM $h\rightarrow \gamma\gamma$
 searches are $|\cos{\theta^*}_{\gamma\gamma}|$~\cite{Mrenna:2000qh} and photon
 balance. The first is the scattering angle of the photons
 calculated in the photon pair rest frame. Cutting on
 $|\cos{\theta^*}|$ has a similar effect to our $\Delta
 \phi_{\gamma\gamma}$ cut, but it is not as efficient. Photon balance
 is defined as the ratio of the leading photon $p_T$ to the sum of
 all photon $p_T$~\cite{unknown:1999fr}. We do not find it to be a useful
 variable for the signal and
 backgrounds we are considering.

%\end{itemize}

% Even with additional angular cuts, the higgs signal is
% barely visible above the background.

Even with additional angular cuts, the Higgs signal in the majority of
parameter space appears only as a slight excess across several bins rather than as a well
defined peak. In fact, to get a Higgs peak at {\em any} ($m_h, m_a$)
point, the branching ratio must be much larger
than in our benchmark value. A peak is first visible for $m_h \sim
150\ \gev$ and $m_a \sim 25\ \gev$ when $BR(h\rightarrow 2\gamma 2g)
\sim 0.04$. Further increasing the branching ratio, the Higgs peak is visible
over a wider region of $m_h, m_a$ space - the lower the Higgs mass, the
larger the branching ratio needs to be before the peak becomes
visible. 

Imposing cuts $\Delta \phi_{jj} < 1.0,\ \Delta
\phi_{\gamma\gamma} < 1.3$, and $\Delta M < 15\ \gev$, the next step
is to determine the significance.
To determine the significance, we count the number of expected signal
and background events in a window $\pm \sqrt 2 \Delta M_{fit}$, where
$\Delta M_{fit}$ is the fitted width of the signal. $10\ \gev$ bins
are used at all times. To make our
estimates more realistic, we include the systematic uncertainty of
$10\%$~\cite{unknown:1999fr, unknown:2006fr} for
the overall jet energy scale and jet resolution. The significance including systematic
uncertainties is given by~\cite{unknown:2006fr, Rainwater:2007cp}
\be
\frac{S}{\sqrt B} \rightarrow \frac{S}{\sqrt{B(1 + \Delta^2 B)}},
\ee
where $\Delta$ is the uncertainty. From this significance we calculate
the branching ratio necessary for discovery given $\CL =
300\ {\rm fb}^{-1}$.  The results are shown below in fig.~(\ref{fig:gothdp300}).

% To estimate whether or not the Higgs peak is visible at a particular
% $m_h, m_a$ point with a given luminosity and branching ratio, we compare the
% combined signal plus background $(S + B)$ distribution to the background with a
% $+10\%$ fluctuation, $(1.1\times B)$. We calculate the $\chi^2$ of
% both $S+B$, and $1.1\times B$ relative to the background fit and keep only
% the signals with  $\chi^2_{S+B} > \chi^2_{1.1B}$. For signals that pass this test, we
% fit the combined $S + B$ to a gaussian plus background, then
% determine the significance by counting signal and background events within
% $\pm {\rm max} \{25\ \gev, 2\Delta M_{fit}\}$, where $\Delta M_{fit}$ is the
% fitted signal width. We repeat this process several times, varying the
% most important backgrounds according to their uncertainty, and use the
% average values.

% For signals with an obvious peak at a given branching ratio, we
% fit the combined $S + B$ to a gaussian plus background, then
% determine the significance by counting signal and background events within
% $\pm {\rm max} \{25\ \gev, 2\Delta M_{fit}\}$, where $\Delta M_{fit}$ is the
% fitted signal width. 

% BETTER TEST STAT. SIGNIFICANCE OF THIS PROCEDURE..

\begin{figure}[!h]
  \includegraphics[width=2.7in,height=2.5in]{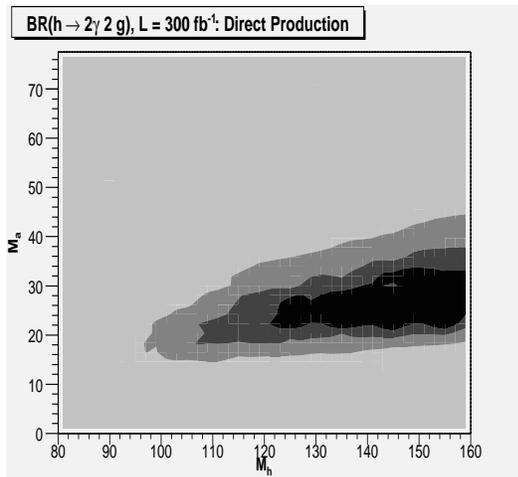}
  \caption{Branching ratio consistent with $S/\sqrt{B} \ge 5.0$ for
    direct Higgs production with $\CL =
    300\ {\rm fb}^{-1}$. The contours indicate discovery branching ratios of
    $\simle 0.06\ {\rm (darkest\ region) },0.1, 0.2\ ({\rm
      lightest})$. Heavy Higgses pass the isolation and $p_T$ cuts
    more easily and are therefore visible at smaller branching
    ratios. The $m_a$ range is restricted above by the $\Delta \phi$
    cuts, and below by the isolation cut. We do not consider branching
  ratios larger than $0.2$.}
  \label{fig:gothdp300}
\end{figure}

As expected, a Higgs with the benchmark branching ratio
is too small to be detected, even with $\CL = 300\ {\rm fb}^{-1}$. Higgses with
branching ratio $\sim 0.06$ are visible in a window around $m_h \sim 140\ \gev$, $m_a \sim 20\
\gev$, but the size of the detection window doesn't increase very quickly
with increasing branching
ratio. Notice that a Higgs lighter than $90\
\gev$ require $BR(h\rightarrow 2\gamma 2g) > 0.2$. The limited $m_a$
extension of the window is a result of the $\Delta \phi$ and isolation
cuts. A
typical Higgs mass resolution in the discovered region is $\sim\ 10\ \gev$.

We used the above procedure in order to get a rough estimates the
significance throughout the $m_h, m_a$ parameter space. At any
particular $(m_h, m_a)$ point, we expect the significance
can be improved by further optimizing the cuts and using more
sophisticated significance techniques. Another way to improve the
significance in this channel is to perform a simultaneous
search for the pseudoscalar in $M_{\gamma\gamma}$. Once the
pseudoscalar is discovered, the small
width of the pseudoscalar will allow us to eliminate a lot of
background. Rather than pursue this here, we will move on to a
different, cleaner production mode.

\section{Associated Production:}

The second production mechanism we consider is associated production,
$pp \rightarrow W^{\pm}h$. Because we use the hadronic decays of the pseudoscalar,
$a\rightarrow jj$, we cannot use the hadronic decays of $W^{\pm}$. The signal is therefore $\ell
\slashchar{E_T} + \gamma \gamma jj$, $\ell = e, \mu$ which we simulate
using PYTHIA.  We assume a Standard Model production cross section, and
scale our PYTHIA-level cross section to match to match the values in
Ref.~\cite{Djouadi:2005gi}. The events are passed through PGS, where
leptons are accepted by if they have $p_T > 5.0\ \gev$, if the $p_T$
of all tracks within $\Delta R = 0.4$ is less than $5.0\ \gev$, and if the $E_T$ in $3\times 3$ calorimeter segment collar around the lepton is less that
$1.1\times$ the $E_T$ of the lepton.

The $pp\rightarrow W^{\pm}h$ production cross section at the LHC is smaller than $pp\rightarrow h$ by a factor of $\sim 20$
at $m_h \sim 100\ \gev$, and it falls off faster with increasing
$m_h$~\cite{Djouadi:2005gi}. The smaller cross section, combined with the small
branching ratio, means we will have far fewer events at a given luminosity. Because $\sigma(pp\rightarrow Z^0 h) <
\sigma(pp\rightarrow W^{\pm}h)$ and $BR(Z^0 \rightarrow \ell^+ \ell^-)
< BR(W \rightarrow \ell \nu)$, we neglect associated $Z^0h$
production.

In addition to the cuts for direct production, we impose:
\begin{itemize}
\item $1$ lepton with $p_T > 20\ \gev, |\eta| < 2.5$
\item $\Delta R_{\ell-\gamma}, \Delta R_{\ell-j} > 0.4$
\item $\slashchar{E}_T > 20.0\ \gev$
  % \item $\Delta R_{\ell-\slashchar{E}_T} > 0.7$ 
\end{itemize}

% The $\slashchar{E}_T$ cut could be hardened to further suppress
% the backgrounds, but it is not necessary.

The signal efficiency after the basic cuts varies from $3\%$ to $15 \%$ and depends primarily on $m_a$. It is highest
near $m_a \sim 0.5~m_h$.

%  As with direct production, increasing the jet
% isolation cut to $\Delta R_{jj} >
% 0.7$ is detrimental to the signal. SIZE?

\subsection{Background}

Fortunately, the background is much smaller than it was for direct production. The lepton and
missing energy are useful for reducing the QCD background, and
$W + N\ \jets$ is suppressed by the small jet-$\gamma$ fake rate. The primary
background is $W + m\ \gamma + N\ \jets$, where $m = 1,2$ is
the number of photons produced with the $W$. 

The diphoton background $\gamma\gamma + QCD$, where a jet fakes a lepton, can
be sizable if the $\slashchar{E}_T$ cut is low. The lepton fake
rate is small $\CO(10^{-4})$~\cite{:1999fq}, but it isn't enough to completely
suppress the large QCD cross sections. With the current 
cuts we find it constitutes less than $1\%$ of the background. Harder
$\slashchar{E}_T$ cuts or a cut on
the transverse $W$ mass can be imposed to suppress it further.

Another potential background is $\tau^+ \tau^-$ production, 
where one $\tau$ decays leptonically. These events have a lepton and contain missing energy from the
neutrinos in both $\tau$ decays. Photons arise
in $\tau$ decay through rare decay modes like
$\tau\rightarrow e^-\bar{\nu}_e \nu_{\tau}\gamma$, through final
state radiation, and through hadronic $\tau$ jets faking
photons. We estimate that isolation cuts, combined with small branching
ratios and the small jet-fake rate, render this background small.
 We do not include it in the full analysis for simplicity.

The complete set of backgrounds is:
\begin{itemize}
\item $W\gamma\gamma + N\ \jets,\ N = 0,1,2$
\item $W\gamma + \jets + N\ \jets,\ N = 0,1,2$
\item $W + N \jets, N = 2, 3, 4$
%\item $\tau^+ \tau^- + \jets\ ( + \gamma s)$
\item ${\rm diphoton + fake\ lepton}$
\end{itemize}

The $W + \jets$ and $W +
m\ \gamma + N\ \jets$ backgrounds were simulated using ALPGEN with factorization scale
$Q^2_{fac} = m^2_W + p^2_{T,W}$. The $K$ factor for $W(\rightarrow
\ell \nu) + 2\ \jets$ is
$\approx 1$~\cite{Oleari:2003tc, Campbell:2003hd}, which we assume for
$W(\rightarrow \ell \nu) + \jets + m\ \gamma $ as well.
The diphoton events generated for
direct production are used again here with a fake lepton
rate $1.25\times 10^{-4}$. As before, all background events were
hadronized in PYTHIA and run through PGS. 

%  A cut
% on the $W$ transverse mass can suppress it further if
% necessary. 

% The cut on $\Delta
% R_{\ell \slashchar{E}_T}$ was included to get the ALPGEN multijet
% cross sections to converge faster ($\sigma(W + 4\jet) \ll \sigma(W
% + 3\jet) \ll \sigma(W + 2\jet)$.. ). CHECK... UNNECESSARY.

% Explicitly generating $W + \gamma\gamma + 2^+ \jets$ in MADGRAPH is very slow
% and difficult. Instead, we calculate $\sigma(W + \gamma\gamma + \jet)$ then use values for  $\frac{W +
%   2^+\jets}{W + 1 \jet}$ (calculated in MG, ALP,..) to estimate
% $\sigma(W +\gamma\gamma + N\ \jets)$. This basically results in an
% enhancement of $\sim 2.5$, which is consistent with the value
% we get for $\frac{\gamma\gamma + 2^+ \jets}{\gamma\gamma +
%   \jet}$. However, since the kinematics (and efficiency) of the
% higher jet events may be slightly different, an analysis with fully
% simulated $W + \gamma\gamma + N\ \jet$ events would be best,
% especially because we want to use kinematics to distinguish signal from
% background. MADGRAPH uses fixed renormalization and factorization scale, which we take
% to be $q_{fac} = 400\ \gev$.

% Varying $q_{fac}$ by a factor of $\CO(2)$ changes the cross section by only $\CO(10\%)$.

\subsection{Reconstruction}

% Despite the smaller cross section, the background for associated
% production are small enough that the Higgs can be discovered without a
% simultaneous pseudoscalar search. A simultaneous search could be done in
% this channel as well and would obviously increase the
% significance, however a single search is much simpler and requires
% fewer assumptions about the background.

Combining the two photons with the leading two jets, we form the four body invariant mass
$M_{jj\gamma\gamma}$. The signal can be enhanced by imposing $\Delta \phi_{jj}$ and $\Delta
\phi_{\gamma\gamma}$ cuts. As in direct production, these cuts are
most effective for a light pseudoscalar. It is also helpful to impose the $\Delta M \equiv
|m_{jj} - m_{\gamma\gamma}| \le 20\ \gev$ cut.
Because there are fewer signal events, we must
be careful that the additional cuts to not limit us to $ < 5$
events (at a given luminosity) necessary for discovery.

%  where the $\Delta \phi$
% distributions peak near $\Delta \phi \approx 0.5$.

% Fitting the combined signal plus background to a gaussian peak on top
% of the background, we find the Higgs mass resolution is typically $\sim
% 8\ \gev$. The width does change with $m_a$ and $m_h$, so to be
% conservative we estimate the significance is by counting signal and
% background events with $\pm 20\ \gev$ of the signal peak using
% $10\ \gev$ bins.

After applying angular cuts
of $\Delta \phi_{\gamma\gamma} \le
1.5, \Delta \phi_{jj} \le 1.3\ {\rm and}\ \Delta M \le 15\ \gev$, we
estimate the significance using the same method as in direct
production. The Higgs mass resolution is typically $\sim 8-10\ \gev$, and we again
include a $10\%$ systematic jet energy uncertainty. Using this significance, we
calculate the branching ratio
required for discovery (significance $>5$) at $\CL = 300\ {\rm
  fb}^{-1}$ as a function of $m_h$ and $m_a$. It is plotted below in fig.~(\ref{fig:gothap300}).

\begin{figure}[!h]
  \includegraphics[width=2.7in,height=2.5in]{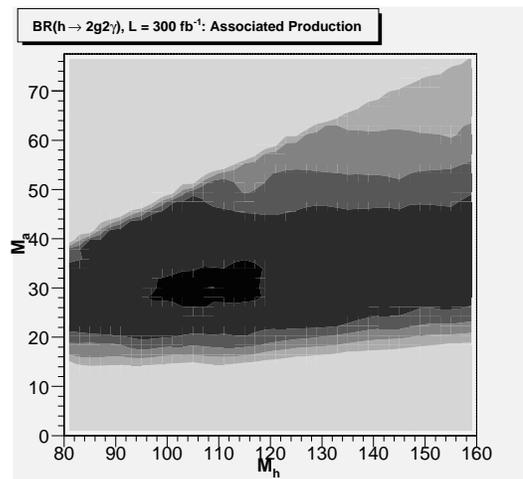}
  \caption{Branching ratio consistent with $S/\sqrt{B} \ge 5.0$ in
    associated Higgs production with $\CL =
    300\ {\rm fb}^{-1}$ of luminosity. We require at least $5$ events for
    discovery. Contours in indicate branching ratios of:
    $\simle 0.02 ({\rm darkest\ region}),0.04, 0.06, 0.1,0.2$~(lightest). The photon
    reconstruction and cut efficiency increases with $m_h$, offsetting
  the decreasing production cross section. }
  \label{fig:gothap300}
\end{figure}

For $m_a$ between $20\ \gev$ and $45\ \gev$, the branching ratio
$BR(h\rightarrow 2\gamma 2g) = 0.04$ is sufficient for Higgs discovery for any $m_h < 160\ \gev$
value. The remaining parameter space would be explored with relatively
small $\sim \CO(2)$ increases in the
branching ratio (or luminosity).  Within the narrow range $100\ \gev < m_h < 120\ \gev$
and $m_a \sim 30\ \gev$, a Higgs with even smaller branching ratio would be
discovered. 

The cut values $\Delta \phi < 1.5,\ 
\Delta \phi_{jj} < 1.3$ were chosen to yield high significance for light
Higgs masses. The angular cuts are less severe than in direct
production, which increases
the $m_a$ reach of our search. Keeping $m_a$ constant and increasing
$m_h$, we see that the production cross section decreases but the cut efficiency
increases. The result is a nearly constant significance.
Loosening the $\Delta \phi$ cuts allows us to better probe the
$m_a > 45\ \gev$ region, however more
sophisticated significance estimates are necessary there because the
signal becomes broad.

Simply comparing fig.~(\ref{fig:gothap300}) and the $h\rightarrow 4\gamma$
mode~\cite{Chang:2006bw} using Eq.~(\ref{eq:compare}), shows both
modes have similar sensitivity for $m_h \lessapprox 100\ \gev$ while
$h\rightarrow 4\gamma$ is more sensitive at larger $m_h$. However, this simple comparison somewhat misleading because
analysis in Ref.~\cite{Chang:2006bw} was done without a detector
simulator. While $h\rightarrow 4\gamma$ will be immune to most
complications of a detector simulator, it is sensitive to the photon reconstruction
efficiency. We find the photon efficiency using the PGS reconstruction
procedure described in section III. to range from $65 - 80\%$, depending on $m_h$ and $m_a$. In comparison,
Ref.~\cite{Chang:2006bw} used an efficiency of $80\%$. Put on equal footing efficiency-wise, we expect the $h\rightarrow 4\gamma$ and
$h\rightarrow 2\gamma 2 g$ modes to be comparably sensitive over a
wider range of $m_h,m_a$ values and to complement each other well.
Because the background in the $h\rightarrow 4\gamma$ mode is so small,
it is sensitive to the large-$m_h$, large-$m_a$ region - exactly the
region that is missed in associated production due to the $\Delta
\phi$ cuts. Another important aspect that a simple
comparison misses is the Higgs mass resolution - $\delta m_h \sim 8\ \gev$ in
$h\rightarrow 2\gamma 2g$ while $ \delta m_h \sim 1-2\ \gev$ in $h\rightarrow
4\gamma$.

\section{Conclusions:}

 Electroweak singlets are easy to add to any model of
  beyond-the-SM physics, and they are even
  required in some cases. Their presence
  can cause large deviations from SM Higgs decay patterns.
 Cascade decays $h\rightarrow aa\rightarrow 4X, X\ne b, \bar b$
 can naturally dominate in extra-singlet scenarios and will be missed by
 conventional detection techniques. It is therefore important to
 investigate where and how the LHC should look to discover this type
 of Higgs. 

 The combination of electroweak singlets with two other
 common new physics features, a $\CZ_2$
 symmetry and new vector-like fermions, can lead to a Higgs that is
 particularly difficult to find -- Higgses which decay predominantly to gluons and photons
 and can be as light as $82\ \gev$.  In this paper we have explored
 the discovery potential at the LHC for these elusive Higgses using the
 cascade decay mode $h\rightarrow aa\rightarrow 2\gamma 2g$. We
 explored the search criteria and the corresponding discovery regions in
 $m_h, m_a$ space using a benchmark luminosity of $\CL = 300\ {\rm
   fb}^{-1}$ in both direct and associated Higgs production. We generated events using PYTHIA and ALPGEN, and used
 the detector simulator PGS.
 
%and $BR(h\rightarrow 2\gamma 2g) = 7.6\times 10^{-3}$.
%  Although the benchmark branching ratio $BR(h\rightarrow
%  2\gamma 2g) = 0.0076$ remains out of reach, 

Of the two modes, we find associated production $pp \rightarrow W^{\pm}h\rightarrow \ell
 \nu + 2\gamma 2g$ is sensitive to a wider range of $m_h, m_a$ masses
 at lower $h\rightarrow 2\gamma 2g$ branching ratio. Imposing cuts on
 the azimuthal separation of the pseudoscalar decay products, $\Delta
 \phi_{\gamma\gamma} < 1.5, \Delta \phi_{jj} < 1.3$  and on the
 difference between reconstructed pseudoscalar masses $|m_{jj} -
 m_{\gamma\gamma}| < 15\ \gev$, we maximize the significance for a light
 Higgs. Given $300\ {\rm fb}^{-1}$ and $BR(h\rightarrow 2\gamma 2g) \cong
 0.04$, we find a $\ge 5\ \sigma$ Higgs signal for $20\ \gev < m_a
 < 45\ \gev$. Isolation cuts prevent us from seeing lighter
 pseudoscalars, while at larger $m_a$ the signal becomes too similar
 to the background. 

 Assuming the Higgs always decays to pseudoscalars, associated
 production is comparable to $h\rightarrow 4\gamma$ for $m_h \simle
 110\ \gev$. Combining both modes will
 yield greater significance at a given luminosity.

%  For $m_h = 90 - 120\ \gev , m_a \cong 25\ \gev$ a Higgs with even
%  smaller branching ratio would be discovered.

%  Associated
%  production nicely complements the $h\rightarrow 4\gamma$
%  mode which is most significant
%  at large $m_h, m_a$~\cite{Chang:2006bw}. 
 
 The other mode we considered, direct production $pp \rightarrow h \rightarrow 2\gamma 2g$, has
 a larger signal but much larger background. Imposing angular cuts 
 eliminates some of the background, but forces us to consider a smaller
 pseudoscalar mass range. Even with strict angular cuts, only a small 
slice of parameter space ($m_h \simge 120\ \gev, m_a \sim 25\ \gev$) would be discovered with the
 branching ratios of interest.

% Other mode: What
% about other production mechanisms, $tth, {\rm VBF}, ..$? Combine
%   these modes with $pp\rightarrow h$ to improve
%   significance.
 
\section*{Acknowledgments}
 
We thank Kevin Black for his advice and help throughout this work. We are also grateful to Thomas Appelquist
for useful discussions and for careful reading of drafts. This work is supported by the
U.S. Department of Energy under grant DE-FG02-92ER-40704.

 \bibliographystyle{utcaps}
\bibliography{SLH}

\end{document}